\documentclass[conference,a4paper,pdftex]{IEEEtran}
\usepackage{xcolor}
\usepackage{balance}
\usepackage{cite}
\usepackage[most]{tcolorbox}
\usepackage{subcaption}

\usepackage{subcaption}
\usepackage{algorithm}
\usepackage{algpseudocode}
\usepackage{graphicx} 
\usepackage{footnote}
\usepackage{amsmath}
\usepackage{amssymb}
\usepackage{verbatim}
\usepackage[letterpaper, left=0.625in, right=0.625in, top=0.75in, bottom=1in]{geometry}
\usepackage{xcolor}
\usepackage{soul}
\usepackage{bm}
\usepackage{array} 
\begin{document}
%
\title{


Resource allocation in Distributed ISAC Systems: A Location Sensitive Mode Selection and Task Assignment Framework 

}

\author{\IEEEauthorblockN{Kwadwo Mensah Obeng Afrane, André B.J. Kokkeler, and Yang Miao}                                    
\IEEEauthorblockA{\IEEEauthorrefmark{1}Faculty of EEMCS, University of Twente,
Enschede, The Netherlands \\ 
k.m.obengafrane@utwente.nl,  a.b.j.kokkeler@utwente.nl, y.miao@utwente.nl}}



\maketitle
\begin{abstract}
In this work we propose a region of interest (ROI)-aware mode selection and task assignment (RAMSTA) framework for distributed integrated sensing and communication  (ISAC) systems. In realistic ISAC use-cases, the sensing performance requirement can be location dependent. Vehicular targets at a traffic intersection, unmanned aerial vehicle approaching a restricted airspace or a sparsely populated areas may require different levels of detection or localization accuracy. This warrants the design of location sensitive resource allocation to account for the varying sensing performance requirements. Thus, we design a sigmoid-based ROI proximity parameter to tune the communication-sensing trade-off in a weighted sum rate and position posterior Cramér-Rao lower bound minimization problem. The resulting mixed-integer non-linear program is solved by applying a penalized convex-concave procedure. We show that the ROI sensitivity thresholds allows the adaptive variation of the communication-sensing trade-off with respect to the predicted location of the target and relative to a predefined ROI.

\end{abstract}

\vskip0.5\baselineskip

\section{Introduction}
Integrated sensing and communication (ISAC), a key enabler of next generation wireless networks, enables the joint design of communication and sensing systems by leveraging their similarities in terms of signal processing algorithms and, to some extent, system architecture \cite{9540344}. 
A distributed multiple-input multiple-output (MIMO) ISAC implementation further benefits from improved diversity due to multi-static sensing as well as joint transmit and receive beamforming gain, thus improving sensing performance under sufficient synchronization \cite{multistatic}. 

Mode selection, subcarrier, and power allocation have been widely studied in distributed MIMO ISAC systems \cite{resource,10571110,mode3}.
Specifically, \cite{resource} addresses the communication–sensing tradeoff via joint access point (AP) mode selection, power and subcarrier allocation. This was achieved via a communication rate maximization subject to Cramér-Rao lower bound constraints. Also, \cite{10571110} 
jointly optimize beamforming and mode selection to maximize the sensing SINR under communication SINR constraints. 
Similarly, \cite{mode3} formulates a mixed-integer problem that jointly optimizes mode selection and power to maximize the minimum sensing SINR under communication constraints.

However, some ISAC use-cases, require robustness to time-varying performance constraints. This is especially relevant in target tracking applications such as traffic throughput and safety at road intersections, or safe and economic unmanned aerial vehicle transport. 
Along a target's trajectory, sensing requirements may be stricter in certain regions of interest (ROIs), such as dangerous intersections or restricted airspace near airports, where more accurate tracking and localization accuracy is required.
The above works do not account for such a ROI-aware system design and to the best of our knowledge this problem has not been studied. Thus, we propose an ROI-aware mode selection and task assignment (RAMSTA) framework  which allows the adaptive variation in communication-sensing trade-off with respect to the predicted location of the target relative to the ROI. Our contributions are enumerated as follows:
\begin{itemize}
    \item We propose a RAMSTA framework for distributed MIMO ISAC systems, which jointly assigns transmit or receive operation modes, communication, sensing or ISAC tasks using a sigmoid-based ROI proximity parameter to adaptively tune the communication-sensing trade-off.
    \item We derive the posterior Cramér Rao Lower bound (PCRLB) for position tracking and minimize the weighted sum of the position PCRLB and sum-rate with mode selection and task assignment as optimization variables. 
    \item The resulting mixed-integer non-linear program (MINLP) optimization problem is subsequently solved  by applying a penalized convex-concave procedure (PCCP). 
    \item We further employ an ROI-aware heuristic subcarrier allocation and an ROI-aware non-uniform power allocation.
\end{itemize}

\section{SYSTEM MODEL}
\begin{figure}[tb]
\centering
\includegraphics[width=0.25\textwidth]{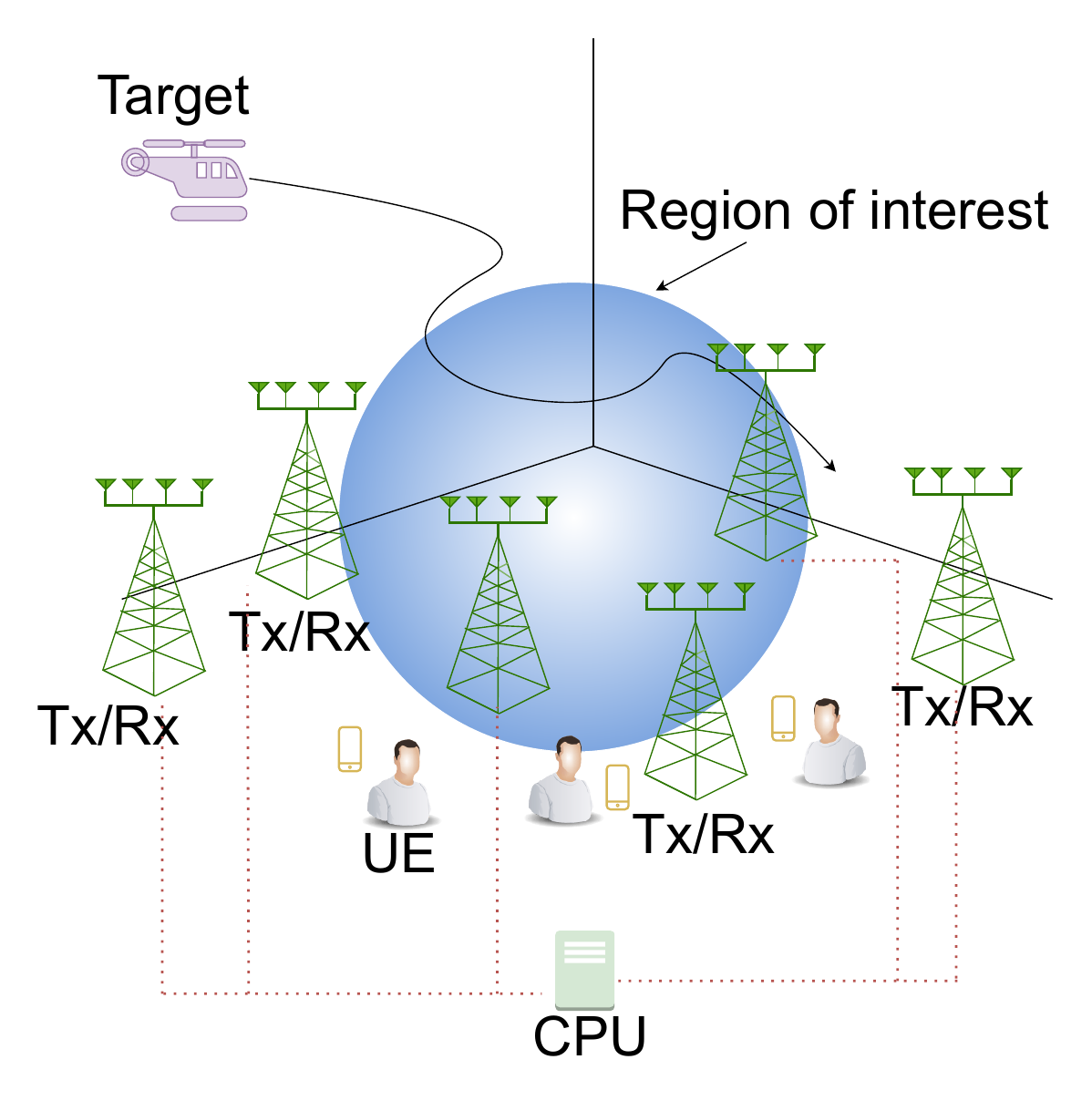}
\caption{\footnotesize{System Model} }
\label{fig1}
\end{figure}

We consider a distributed MIMO system Fig. \ref{fig1}, where $J$ APs serve $K$ single antenna downlink user equipment (UEs) and simultaneously track a single target. Each AP is equipped with a uniform planar array with $L_{t} = L_{t}^x \times L_{t}^z $ and $L_{r} = L_{r}^x \times L_{r}^z $ transmit and receive antenna elements respectively. The ROI is characterized by a sphere with a known center $\bar{\mathbf{c}}$ and radius $\bar{r}$ within which a higher quality of sensing service is required. 

In mode selection and task assignment, the objective is to jointly assign transmit or receive operation modes, communication, sensing or ISAC tasks to APs.
For each tracking interval $i$, the mode selection variable of the $j$-th AP is $a_j^ i \in \{0,1 \}$, where $a_j^ i=1$ or $a_j^ i=0$ indicates a transmit or receive mode respectively. 
The task assignment vector of the $j$-th transmit AP is $\bar{\mathbf{a}}_j^i = [\bar{a}_{j,0}^i, \bar{a}_{j,1}^i, \ldots, \bar{a}_{j,K}^i]$, where $\bar{a}_{j,0}^i=1$ indicates a sensing task and $\bar{a}_{j,k}^i=1$, $k=1,\ldots,K$, indicates DL communication with the $k$-th UE.

\subsection{Sensing Model}
\subsubsection{Signal Model}
Let the transmitted baseband signal by the $j$-th AP on the $m$-th subcarrier and $n$-th OFDM symbol at the $i$-th tracking interval be
\begin{align}
    \label{transmit}
    \nonumber
    \mathbf{s}_{j}^i(m,n) &= a_{j}^i\Big[
        \sum_{k=1}^K\bar{a}_{j,k}^ i\sqrt{p_{j,k}(m)}\mathbf{w}_{j,k}^i\eta_k^i(m)\bar{s}_{j,k}^i(m,n) 
    \\
    &+ \bar{a}_{j,0}^ i\sqrt{p_{j}^i(m)}\mathbf{w}_{j}^i\eta_{j}^i(m)\bar{s}_{j}^i(m,n)\Big],
\end{align}
where $ p_{j,k}(m)$, $\eta_k^i(m)\in \{0,1\}$, $\mathbf{w}_{j,k}^i \in \mathbb{C}^ {L_t\times 1}$ ,  $\bar{s}_{j,k}^i(m,n)$, are the transmit power, subcarrier allocation variable, beamforming vector and communication symbol for the $k$-th DL UE respectively. Also,  $\mathbf{w}_{j}^i \in \mathbb{C}^ {L_t\times 1}$, $\eta_{j}^i(m)\in \{0,1\}$, $ p_{j}^i(m)$, $\bar{s}_{j}^i(m,n)$ characterize the sensing beamformer, sensing subcarrier allocation variable, sensing power allocated to the $j$-th AP on the $m$-th  subcarrier, and the baseband sensing symbol respectively. 

The DL APs transmit a probing signal to illuminate the target, and the UL APs receive the reflected echoes.
The received demodulated OFDM signal on the $m$-th subcarrier and $n$-th OFDM symbol of the $r$-th UL AP from the $j$-th DL AP at the central processing unit (CPU) is;
\begin{align} 
\label{rad}
\nonumber
    \mathbf{y}_{r,j}^i (m,n) &= \mathbf{G}_{r,j}^i \mathbf{w}_{j}^i\bar{a}_{j,0}^i(1-a_r^i)a_j^i\eta_{j}^i(m)\sqrt{p_{j}^i(m)}\bar{s}_j^i(m,n)  \\
    &+  \mathbf{Q}_{r,j}^i(m) \mathbf{s}_{j}^i(m,n) + \mathbf{n}_r
\end{align}
where $\mathbf{G}_{r,j}^i \in \mathbb{C}^ {L_r\times L_t}$, $\mathbf{Q}^i_{r,j}(m) \sim \mathcal{CN}(0,\sigma^2_Q\mathbf{I}_{L_r \times L_t})$, and $\mathbf{n}_r\sim \mathcal{CN}(0,\sigma_{r}^2\mathbf{I}_{L_r})$ are the bistatic radar channel (\ref{rad_channel}), AP-AP channel estimation error matrix \cite{resource} 
, and noise respectively. 
\begin{equation}
    \mathbf{G}_{r,j}^i=\sqrt{\alpha_{j,r}}e^{\mathrm{j}2\pi(nT_sf_{r,j}^i-m\tau_{r,j}^i\Delta f)}\mathbf{a}_{r}(\theta_{r}^i,\phi_{r}^i)\mathbf{a}_{t}^H(\theta_{j}^i,\phi_{j}^i)
    \label{rad_channel}
\end{equation}
In (\ref{rad_channel}), $\alpha_{r,j} $, $T_s$, $f_{r,j}^i$, $\tau_{r,j}^i$, and $\Delta f$ are the attenuation factor which incorporates the bistatic radar range equation \cite{rad_range}, the OFDM symbol duration including cyclic prefix, Doppler shift, delay, and subcarrier spacing respectively. Also, $\mathbf{a}_{r}(\theta_{r}^i,\phi_{r}^i) \in \mathbb{C}^{L_{r}\times 1}$ and $\mathbf{a}_{t}(\theta_{j}^i,\phi_{j}^i) \in \mathbb{C}^{L_{t}\times 1}$ are the transmit and receive steering vectors of the angle of arrival (AoA) and angle of departure (AoD) in the horizontal ($\theta$) and vertical ($\phi$) directions of the target respectively. $\mathbf{a}_{t}(\theta_{j}^i,\phi_{j}^i) =\mathbf{a}_{t}^x(\theta_{j}^i,\phi_{j}^i)\otimes \mathbf{a}_{t}^z(\phi_{j}^i)$ \cite{steer} where, $    \mathbf{a}_{t}^x(\theta_{j}^i,\phi_{j}^i) = [1, e^{\mathrm{j}2\pi d_x \Psi_j^i/\lambda },\cdots, e^{\mathrm{j}2\pi d_x \Psi_j^i (L_t^x-1)/\lambda }]^T$,  $   \mathbf{a}_{t}^z(\phi_{j}^i) = [1, e^{\mathrm{j}2\pi d_z \Omega_j^i/\lambda },\cdots, e^{\mathrm{j}2\pi d_z \Omega_j^i (L_t^z-1)/\lambda }]^T$, 
and $\otimes $ is the Kronecker product. Here, $\lambda$, $d_x$, and $d_z$ are the wavelength, antenna spacing along $x$ and $z$ axes respectively. Whereas $\Psi_j= \cos(\phi_j^i)\cos(\theta_j^i)$ and $\Omega_j^i=\sin(\phi_j ^i)$. For simplicity, we use a conjugate sensing beamformer in the predicted AoD direction of the target, $\mathbf{w}_{j}^i = \mathbf{a}_{t}(\theta_{j}^{i|i-1},\phi_{j}^{i|i-1})/\sqrt{L_t}$.

\subsubsection{Target Tracking Model} The target motion between tracking intervals, $i$ and $i-1$ follows a nearly constant velocity model \cite{motion_model}, $\mathbf{x}^i =\mathbf{F}\mathbf{x}^{i-1} + \mathbf{e}^{i-1}$
where $\mathbf{x}^{i} = [x^i, y^i, z^i, \dot{x}^i, \dot{y}^i, \dot{z}^i]^T$ is the target state. $(x^i, y^i, z^i)$ and $(\dot{x}^i, \dot{y}^i, \dot{z}^i)$ denote the 3D position and velocity respectively. $\mathbf{F}=\begin{bmatrix}
1 & \Delta T  \\
0 & 1 \\
\end{bmatrix} \otimes \mathbf{I}_3$ is the state transition matrix
where $\Delta T$ and $\mathbf{I}_3$ are the time difference between adjacent tracking intervals and a $3 \times 3$ identity matrix respectively. $\mathbf{e}^{i-1}  \sim  \mathcal{N} (0,\mathbf{E}^{i-1})$ is the process noise where, $    \mathbf{E}^{i-1} = \bar{\sigma}\begin{bmatrix}
\Delta T^3/3 & \Delta T^2/2 \\
\Delta T^2/2 & \Delta T 
\end{bmatrix}  \otimes \mathbf{I}_3,$.
$\bar{\sigma}$ is the noise power spectral density . The measurement function of the $(r,j)$-th bistatic pair at the $i$-th tracking interval is 
\begin{equation}
    \label{meas_func}
    \mathbf{z}_{r,j}^{i} = \mathbf{h}_{r,j}(\mathbf{x}^{i}) + \tilde{\mathbf{e}}_{r,j}^{i},
\end{equation}
where $\mathbf{h}_{r,j}(\mathbf{x}^{i})= [f_{r,j}, \tau_{r,j}^{i},\theta_{r}^{i}, \phi_{r}^{i}]^T $ is the measurement function whereas $\tilde{\mathbf{e}}_{r,j}^{i} \sim \mathcal{N}(0,\mathbf{R}_{r,j}^i)$ is the measurement noise whose covariance $\mathbf{R}_{r,j}^i$ is derrived in Appendix \ref{sec:AppendixA}.
As the tracking performance metric, we use the PCRLB, whose Fisher information matrix (FIM) is computed recursively as \cite{pcrlb}:
\begin{equation}
\label{pcrlb}
    \mathbf{J}(\mathbf{x}^i) = [\mathbf{E}^{i-1} + \mathbf{F} \mathbf{J}^{-1}(\mathbf{x}^{i-1})\mathbf{F}^T]^{-1} + \mathbf{J}_D
\end{equation}
where $\mathbf{J}_D$, (\ref{data_f}), is the data FIM and is derived from (\ref{rad}). Hereafter, the tracking interval index $i$ is dropped for brevity.
 \begin{equation}
 \label{data_f}
     \mathbf{J}_D =\sum_{m=1}^M\sum_{n=1}^N\sum_{j=1}^J\sum_{r=1}^J\frac{1}{\sigma_{r,j}^2} (1-a_r)a_jp_{j}^i(m)\bar{a}_{j,0}a_j\eta_{j}^i(m)\tilde{\mathbf{J}}(\mathbf{u})
 \end{equation}
 where $\sigma_{r,j}^2 = a_j\sigma_Q^2 \big(\sum_k^K\bar{a}_{j,k}p_{j,k}(m)\eta_k(m) + \bar{a}_{j,0}a_jp_{j}^i(m)\eta_{j}^i(m)\big) + \sigma_r^2$. 
Refer to Appendix \ref{sec:AppendixA}, for proof and definition of $\tilde{\mathbf{J}}(\mathbf{u})$.
The sensing objective used in the optimization is the predicted position PCRLB, $ \mathbb{J} = Tr\Big( \big[(\mathbf{J}(\mathbf{x}^i)|_{\mathbf{x}^{i|i-1}})^{-1} \big]_{1:3,1:3} \Big)$, where (\ref{data_f}) is evaluated at the predicted state of the target \cite{xiong_coalition_2023}.

\subsection{Communication} In the DL transmission, transmit APs cooperatively serve a subset of UEs specified by $\bar{a}_{j,k}$, $\forall k=1,\cdots, K$. Thus, 
the achievable rate of the $k$-th UE is,
\begin{align}
\label{rate}
\nonumber
    R_k &= \sum_{m=1}^M \eta_k(m)\log_2\big( 1 + \gamma_k(m)\big),\\
    \gamma_k(m) &= \frac{\big|\sum_{j=1}^J a_j\bar{a}_{j,k}\mathbf{g}_{j,k}^H(m)\mathbf{w}_{j,k}(m)\sqrt{p_{j,k}(m)}\big|^2}{\sigma^2_{DL}}      
\end{align}
where $\mathbf{g}_{j,k}(m) \in \mathbb{C}^{L_t\times 1}$, $\mathbf{w}_{j,k} = \mathbf{g}_{j,k}(m)/||\mathbf{g}_{j,k}(m)||$, and $\sigma^2_{DL}$  is the DL channel gain and the maximum ratio transmission beamformer, and receiver noise respectively.

\section{Problem Formulation}
\subsection{ROI-aware tuning parameters}
Let $\bar{\mathcal{R}}\triangleq\{\mathbf{x}\in\mathbb{R}^{3}:\|\mathbf{x}-\bar{\mathbf{c}}\|<\bar{r}\}$ and $ d^i \triangleq
    \left\|
    [\hat{\mathbf{x}}^{i|i-1}]_{1:3}
    - \bar{\mathbf{c}}
    \right\|$
denote the ROI and the Euclidean distance between the predicted target position and the ROI center, $\bar{\mathbf{c}}$ respectively. Then we define a sigmoid based ROI proximity parameter, 
\begin{equation}
\label{prox}
    \bar{\rho}^i
    \triangleq
    \operatorname{Sig}(\kappa s^i)
    =
    \frac{1}{1+\exp(-\kappa s^i)},
    \qquad \kappa>0.
\end{equation}
where, $   s^i \triangleq \frac{\bar{r}-d^i}{\bar{r}}$ is the normalized proximity margin. $\bar{\rho}^i \in (0, 1)$ is dimensionless and monotonically decreasing in $d^i$. Thus as the target approaches the ROI, $s^i$ increases and $\bar{\rho}^i$ approaches unity. Thus, $\bar{\rho}^i$  is used to adaptively tune the communication--sensing tradeoff parameter, $\delta^i$, such that
\begin{equation}
    \delta^i
    =
    \delta_{\min}
    +
    (\delta_{\max}-\delta_{\min})\bar{\rho}^i,
    \label{trade_off}
\end{equation}
where $\delta_{\min}$ and $\delta_{\max}$, referred to as ROI sensitivity thresholds, bound $\delta^i$.
In (\ref{trade_off}), $\delta^i$ approaches $\delta_{\max}$ or $\delta_{\min}$ as the target moves closer or further away from the center of the ROI respectively. 

\subsection{RAMSTA}
The joint mode selection and task assignment optimization is formulated as:
\begin{align}
\label{objective}
\mathbf{P1}:\min_{\mathcal{R}} &\quad  \delta^i \beta \mathbb{J} - (1-\delta^i)\sum_{k=1} ^KR_k\\
\label{subcarrier}
\text{s.t.} &\quad \bar{a}_{j,0} + \sum_{k=1} ^K\bar{a}_{j,k} \geq a_j , \quad \forall j \\
  \label{task2}
  & \bar{a}_{j,0} + \sum_{k=1} ^{K}\bar{a}_{j,k} \leq a_j(K +1), \quad \forall j \\
    \label{task3}
  & \sum_{j=1}^Ja_{j,0} \geq 1 \\
  \label{task4}
  & 1 \leq \sum_{j=1}^{J}\bar{a}_{j,k} \leq \upsilon, \quad \forall k\\
  \label{mode}
  & 1 \leq \sum_{j=1}^Ja_j \leq  J-1 \\ 
  \label{bin_task}
  & \bar{a}_{j,0} \in \{0,1\}, \bar{a}_{j,k} \in \{0,1\},  a_{j} \in \{0,1\} ,  \forall j,k 
\end{align}
where, $\mathcal{R}\in \big\{  \bar{a}_{j,0} , \bar{a}_{j,k}, a_{j}\mid \forall j,k\big\}$ and $\beta$ is a normalization factor to keep the communication and sensing objective on the same order of magnitude.
(\ref{subcarrier}), (\ref{task2}), and (\ref{task3}) ensure that at least one task is assigned to each transmit AP, only Tx APs are assigned tasks, and a lower bound on the number of APs assigned a sensing task respectively. Whereas (\ref{task4}) bounds the number of APs assigned to a given UE such that $\upsilon=\big \lfloor   2 + \big((J-1)-2\big)\big(1-\delta^i\big)\big\rfloor$. Furthermore, (\ref{mode}) bounds the total number of transmit APs. 
Power and subcarriers are allocated via non-uniform and heuristic approach respectively, thus are not included as optimization variables or constraints. $\mathbf{P1}$ is a non-convex MINLP problem. 
To obtain a convex reformulation, we first relax the binary constraints into continuous form \cite{bilinear}:
\begin{align}
\label{cont}
    &a_j = [0,1],\bar{a}_{j,0} = [0,1], \bar{a}_{j,k} = [0,1] \quad \forall j,kd \\
    \label{DC}
    &a_j-a_j^2 \leq 0, \bar{a}_{j,0} -\bar{a}_{j,0} ^2 \leq 0, \bar{a}_{j,k}-\bar{a}_{j,k}^2 \leq 0
\end{align}
Since the left side of the inequalities in (\ref{DC}) is a difference of convex function, hence non-convex, we apply a PCCP \cite{bilinear},\cite{resource}.
At each PCCP iteration $(o)$, the concave components of (\ref{DC}) are replaced by their first-order Taylor approximations, and slack variables, $e_j$, $\bar{e}_{j,0}$, and $\bar{e}_{j,k}$ are introduced. This results in constraints (\ref{pccp1})-(\ref{pccp4}), whose violation incurs a penalty (\ref{penalty}).
\begin{align}
\label{pccp1}
     &(a_j^{(o)})^2 + a_j(1-2a_j^{(o)}) \leq e_j, \quad \forall j,kd \\
    \label{pccp2}
     &(\bar{a}_{j,0} ^{(o)})^2 + \bar{a}_{j,0} (1-2\bar{a}_{j,0} ^{(o)}) \leq \bar{e}_{j,0}, \quad \forall j,kd  \\
    \label{pccp3}
     &(\bar{a}_{j,k}^{(o)})^2 + \bar{a}_{j,k}(1-2\bar{a}_{j,k}^{(o)}) \leq \bar{e}_{j,k}, \quad \forall j,kd \\
    \label{pccp4}
     &e_j\geq 0, \bar{e}_{j,0} \geq 0, \bar{e}_{j,k}\geq 0, \quad \forall j,kd\\
     \label{penalty}
     &\mathcal{P} =\zeta^{(o)}\Big(\sum_{j=1}^Je_j + \sum_{k=1}^{K}\sum_{j=1}^J\bar{e}_{j,k} + \sum_{j=1}^J \bar{e}_{j,0} \Big),
\end{align}
where $ \zeta^{(o)}>0$ is the penalty coefficient.
Due to the non-convex coupling between the optimization variables in (\ref{data_f}) and (\ref{rate}). We further transform as follows:
\begin{enumerate}
    \item In (\ref{data_f}), let $\tilde{a}_{j,r} \triangleq (1-a_r)a_j$ be the contribution of the $(r,j)$-th bistatic pair to the FIM such that $\tilde{a}_{j,r} \in [0, 1]$.
    \item By (\ref{subcarrier}) and (\ref{task2}) we can infer that $\bar{a}_{j,0}=1$ and $\bar{a}_{j,k}=1$ implies that $a_j\bar{a}_{j,0}= \bar{a}_{j,0}$ and $a_j\bar{a}_{j,k}= \bar{a}_{j,k}$.
    \item The noise variance in (\ref{data_f}) is replaced by its upper bound \cite{resource},  $\sigma_{j,r}^2 =\sigma_Q^2 P_{\max} + \sigma_r^2$.
    \item The remaining bilinear product of $ \tilde{a}_{j,r}$ and $\bar{a}_{j,0}$ in (\ref{data_f}) is further defined as $q_{j,r} \triangleq \tilde{a}_{j,r}\bar{a}_{j,0}$.
    \item The SNR in (\ref{rate}) is reformulated as $\hat{\gamma}_k(m)=\frac{\sum_{j=1}^J\sum_{j'=1}^J\ddot{a}_{j,j',k}b_{j,k}(m)b_{j',k}^*(m)}{\sigma^2_{DL}}$ 
    where $b_{j,k}(m)=\mathbf{g}_{j,k}^H\mathbf{w}_{j,k}(m)\sqrt{p_{j,k}(m)} $ and $\ddot{a}_{j,j',k} \triangleq \bar{a}_{j,k}\bar{a}_{j',k}$.
    \item Applying the McCormick linearization \cite{mccormick_computability_1976} to $\tilde{a}_{j,r}$, $q_{j,r}$, and $\ddot{a}_{j,j',k}$ gives
    \begin{align}
    \label{a_tilde}
    \nonumber
    &\tilde{a}_{j,r} \geq 0, \tilde{a}_{j,r}\geq a_j-a_r, \tilde{a}_{j,r}\leq 1-a_r , \tilde{a}_{j,r}\leq a_j \\
    \nonumber
    & q_{j,r} \geq 0, q_{j,r} \geq \bar{a}_{j,0} + \tilde{a}_{j,r} - 1, q_{j,r} \leq \bar{a}_{j,0}, q_{j,r} \leq \tilde{a}_{j,r} \\
    \nonumber
    & \ddot{a}_{j,j',k} \geq 0, \ddot{a}_{j,j',k} \geq \bar{a}_{j',k} +\bar{a}_{j,k}-1,  \\
    &\ddot{a}_{j,j',k} \leq \bar{a}_{j',k}, \ddot{a}_{j,j',k} \leq \bar{a}_{j,k}
\end{align}
\item Finally, the position PCRLB term in the objective is reformulated in semidefinite programming form \cite{sdp}.
\end{enumerate}
The resulting convex form, which can be solved with CVX is,
\begin{align}
    \label{objective2}
\mathbf{P1.1}:\min_{\mathcal{R}_{1.1} }\quad & \delta^i \beta Tr \big([\mathbf{D} ]_{1:3,1:3}\big) - (1-\delta^i)\sum_{k=1} ^K\hat{R}_k + \mathcal{P}\\
\text{s.t.} \quad & \begin{bmatrix}
\mathbf{D} & \mathbf{I} \\
\mathbf{I} & \hat{\mathbf{J}}(\mathbf{x}^i)|_{\mathbf{x}^{i|i-1}} 
\end{bmatrix} \succeq 0 \\
&\text{(\ref{subcarrier})}-\text{(\ref{mode})}, \text{(\ref{cont})}, (\ref{pccp1}) - \text{(\ref{pccp4})}, \text{(\ref{a_tilde})}
\end{align}
where $\mathcal{R}_{1.1} \in \big\{\bar{a}_{j,0} , \bar{a}_{j,k}, a_{j},e_j,\bar{e}_{j,0},\bar{e}_{j,k}, \tilde{a}_{j,r}, q_{j,r},\\
\ddot{a}_{j,j'k}, \mathbf{D}  | \forall j,j',r,k \big\}$. $\hat{\mathbf{J}}(\mathbf{x}^i)|_{\mathbf{x}^{i|i-1}} $ and $\hat{R}_k$ is the resulting predicted position PCRLB after steps 1-4 and the sum rate after the SNR reformulation in step 5 respectively. 
A mode selection and task assignment solution to $\mathbf{P1.1} $ is summarized in Algorithm-\ref{alg:cap}. Here, the algorithm is initialized under a feasible uniform subcarrier and power allocation, denoted by $\eta_j^{i,(0)}(m)$, $\eta_k^{i,(0)}(m)$, and $p_{j,k}^{(0)}(m)$, $p_{j}^{i,(0)}(m)$ respectively.

\begin{algorithm}
\caption{Mode Selection and task assignment}\label{alg:cap}
\begin{algorithmic}[1]
\State \textbf{Initialization:} $\phi_j^{i,(0)}$,$\phi_k^{(0)}$, $p_{j,k}^{(0)}(m)$, $\eta_j^{i,(0)}(m)$, $\eta_k^{i,(0)}(m)$ $p_{j}^{i,(0)}(m)$, $\zeta_{\text{max}}$, $\zeta ^{(0)}$, $o=0$, $o_{\max}$, $\Delta=1e-3$, $\varepsilon>1$
\State \textbf{Output} $\mathcal{R}_{1.1}^*$
\Repeat
    \State Obtain optimal solution for $\mathcal{R}_{1.1}^*$ by solving problem
    \State Set $o \leftarrow o+1$
    \State Update $\mathcal{R}_{1.1}^{(o)}  \leftarrow \mathcal{R}_{1.1}^*$
    \State Set $\zeta ^{(o +1)} \leftarrow   \text{min}(\varepsilon \zeta ^{(o)},  \zeta_{\text{max}}) $
\Until $o = o_{\max}$ or $\text{max}\big(f(a_j),f(\bar{a}_{j,0}) , f(a_{j,k}) \big)\leq \Delta$,  $\forall j,kd$, where $f(x) = \text{min}(x, 1-x)$
\end{algorithmic}
\end{algorithm}

\subsection{Power and Subcarrier Allocation}
Here we employ an ROI-proximity-sensitive heuristic subcarrier and non-uniform power allocation. 
After a feasible mode and task assignment solution from Algorithm-\ref{alg:cap}, the total number of subcarriers assigned to the $k$-th UE and sensing  task are $M_{com,k} = \lfloor((1-\delta^i)M)/K \rfloor $ and $M_{sens} = M-\sum_{k=1}^K M_{com,k}$ respectively. The set of $M_{com,k}$ subcarriers assigned to the $k$-th UE is determined according to the average channel gain. $M_{sens}$ is assigned to transmit APs with a sensing task in an interleaved manner. The procedure is enumerated in Algorithm-\ref{alg:cap2}.

\begin{algorithm}[t]
\caption{ROI-aware Heuristic Subcarrier Allocation}
\label{alg:cap2}
\begin{algorithmic}[1]
\State \textbf{Input:} $K$ , $M$, $|\mathbf{g}_k(m)|^2 $, $\bar{a}_{j,k}^*$, $\bar{a}_{j,0}^*$ ,$\delta^i$, $M_{\text{com},k}$, $M_{sens}$
\State \textbf{Output:} $\phi_j^i(m)$, $\phi_k(m)$, $\forall j,kd$

\State Set $\phi_k \gets 0$ and $c_k \gets 0,\ \forall k$
\State Compute average channel gain for all serving APs per UE
\State Sort channel gain in descending order $\forall k$ and $m$
\For{$m=1,\ldots,M$}
    \State Select the UE $k$ with the highest channel gain for
           subcarrier $m$ among UEs satisfying $c_k<M_{\text{com},k}$
    \If{a feasible UE $k$ exists}
        \State $\phi_k(m)\gets 1$; $c_k\gets c_k+1$
    \EndIf
\EndFor
\State Assign $M_{sens}$ to sensing APs in an interleaved manner
\end{algorithmic}
\end{algorithm}
Lastly, each transmit AP is assigned a total power of $P_{j,\max} = P_{\max}/\sum_{j=1}a_j$. Then for each AP with an ISAC task, the total power is divided between UEs and sensing functions such that , $P_{j,com,k} =  (1-\delta^i) P_{j,\max}/\sum_{k=1}\bar{a}_{j,k}$ and $P_{j,sens}= \delta^i P_{j,\max}$ respectively. $p_j^i(m)$ and $p_{j,k}(m)$ are subsequently obtained by equal distribution across assigned subcarriers. Transmit APs with only a communication or sensing task utilize their total power for the assigned task.

\begin{table}
    \centering
    \begin{tabular}{|>{\raggedright\arraybackslash}p{10mm}|l|>{\raggedright\arraybackslash}p{10mm}|>{\raggedright\arraybackslash}p{13.3mm}|>{\raggedright\arraybackslash}p{10mm}|l|} \hline 
        Parameter & Value & Parameter&Value & Parameter&Value\\ \hline 
         $\Delta f$& $30$ kHz& $f_c$&$6.8$ GHz & $\Delta T$& $100$ ms\\ \hline 
         $P_{max}$& $7$ W& $L_{r}, L_{t}$&9 &$\beta$ & $10^6$ \\ \hline 
         $T_s$&  $\frac{5}{ 4\Delta f }$& $N$& 16 &$\kappa$ &7\\ \hline 
         $M$&  $1596$& $\sigma^2_Q$&$-150$ dBm & $\bar{\sigma}$&$10$\\ \hline 
    \end{tabular}
\caption{\footnotesize{Simulation Parameters}}
\label{tab:tab1} 
\end{table}

\section{Results}
\begin{figure*}[!t]
\centering
\subfloat[]{%
    \includegraphics[width=0.32\textwidth]{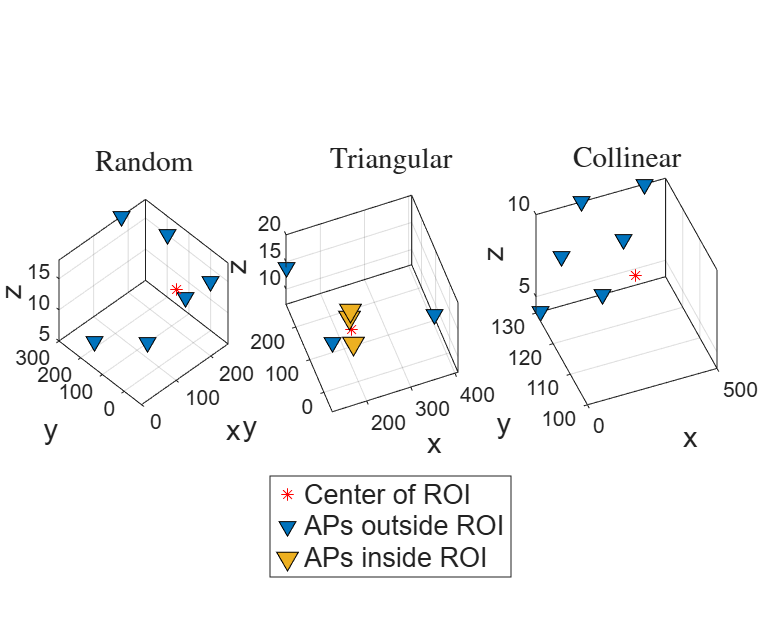}
    \label{fig:second2}
}\hfill
\subfloat[]{%
    \includegraphics[width=0.3\textwidth]
   {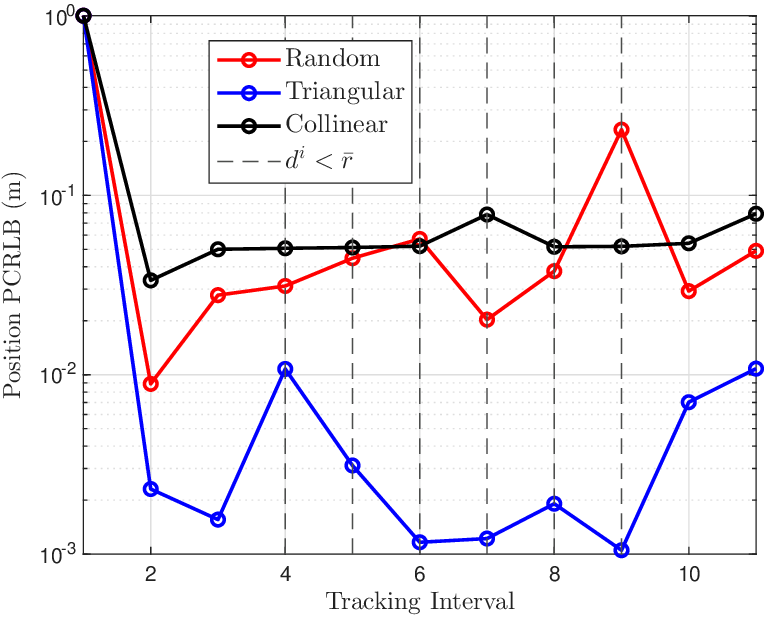}%
    \label{fig:first2}
}\hfill
\subfloat[]{%
    \includegraphics[width=0.3\textwidth]{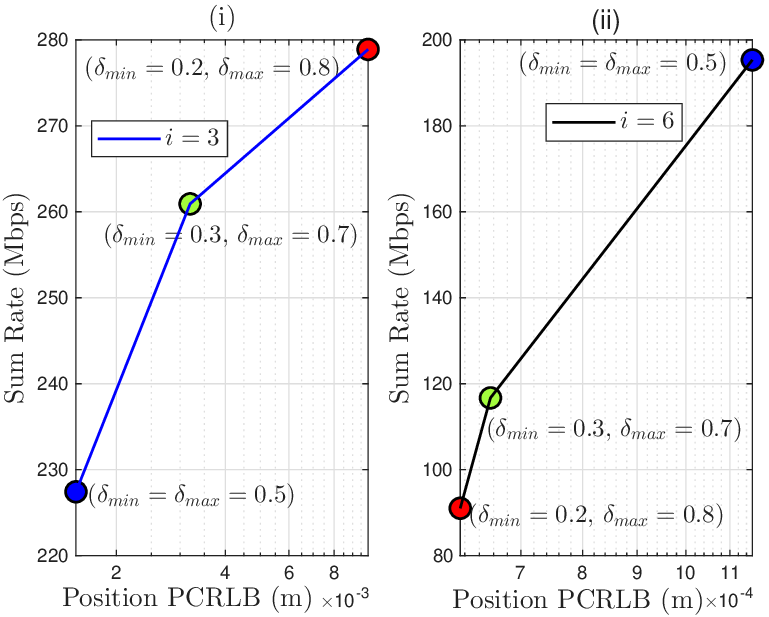}%
    \label{in_out2}
}
\caption{\footnotesize{(a) AP Topology variations relative to ROI (b) Target tracking performance under varying AP topologies ($\delta_{\min}=\delta_{max}=0.5$) (c) Communication-sensing trade-off under varying ROI sensitivity thresholds ( (i) Outside ROI ($i=3$), (ii) Inside ROI ($i=6$) ) }}
\label{fig:three}
\end{figure*}

The simulation setup consists of $6$ APs, $3$ UEs and a single target with a $5$ dBsm radar cross-section. We define an ROI with $\bar{\mathbf{c}} = (217,105,10)$ m, $\bar{r}=7$ m. To obtain an estimate of the state of the target $\hat{\mathbf{x}}^i$, an extended Kalman filter (EKF) is implemented at the CPU where all valid multi-static link  measurements are fused via covariance intersection \cite{multistaticss} \cite{xiong_coalition_2023}. Here $\mathbf{x}^0=[230\text{m},105\text{m},5\text{m},-22\text{m/s},0\text{m/s},7\text{m/s}]^T$. The PCRLB recursion is initialized at $\mathbf{J}(\mathbf{x}^0) = \mathbf{P}^{-1}_0$, where, $\mathbf{P}_0 = \text{diag}\{15\text{m}^2,15\text{m}^2,15\text{m}^2 ,1\text{m}^2/\text{s}^2, 1\text{m}^2/\text{s}^2 1\text{m}^2/\text{s}^2\}$ is the initial covariance used to initialize the EKF. A noise power of $-120$ dBm is assumed whereas, $\mathbf{g}_{j,k}$ is assumed to be Rayleigh distributed with an i.i.d. complex Gaussian small-scale fading vector at zero mean, unit variance, and a large-scale propagation model which follows the 3GPP UMi pathloss with $7.82$ dB shadow fading. Lastly the three non-mobile DL UEs are located at $(140,120,2)$,$(200,150,2)$,$(240,150,2)$ respectively. Lastly, the position PCRLB in meters obtained from $\sqrt{Tr\Big(\big[(\mathbf{J}(\mathbf{x}^i)|_{\mathbf{x}^{i|i-1}})^{-1} \big]_{1:3,1:3} \Big)} $ and the sum rate in Mbps is obtained by dividing total rate by the symbol period ($1/\Delta f$). Other parameters are enumerated in Table-\ref{tab:tab1}.

To determine a suitable ROI sensitive AP deployment, we compare $3$ different topologies;  random, triangular, and collinear in Fig. \ref{fig:second2}. 
The triangular deployment consist of $3$ APs positioned in a triangular formation outside the ROI and $3$ APs inside the ROI whereas in the collinear deployment, pairs of collinear APs are deployed at varying heights. We set  $\delta_{\text{min}}=\delta_{\text{max}} =0.5$ to isolate the impact of topology on performance relative to the ROI. The resulting position PCRLBs are compared in Fig. \ref{fig:first2}. Here, tracking intervals where the target's position falls inside the ROI are denoted as black vertical lines. We observe that topology-b consistently shows a lower position PCRLB inside the ROI. Specifically, the minimal performance difference inside the ROI occurs at $i=4$ where the random, triangular and collinear topologies obtain position PCRLBs of $0.0312 $ m, $0.0108$ m, and $0.0507$ m respectively. Thus, inside the ROI, the triangular topology has a minimum of  $0.0204$ m and $0.0399$ m sensing performance gains against the random and collinear topologies respectively. This gain is largely due to the deployment of APs inside the ROI, forming high SNR multi-static links inside the ROI, whereas the deployment of APs deployed outside the ROI provides diverse observability. Thus, unless otherwise stated the remaining analyses are derived from the triangular topology. 
\begin{figure}[!t]
\centering
\includegraphics[width=0.3\textwidth]{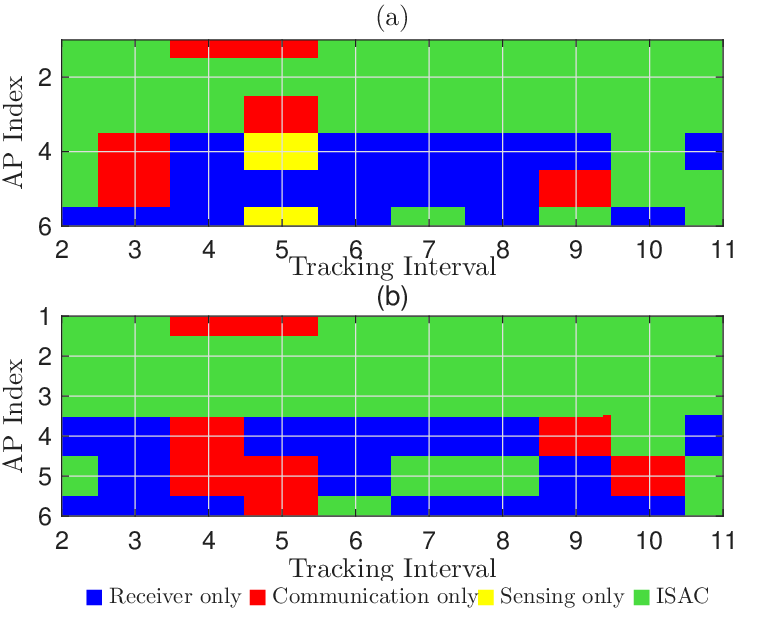}
\caption{\footnotesize{AP mode and task assignment results (a)  $\delta_{min}=0.2$, $\delta_{max}=0.8$  (b) $\delta_{min}=\delta_{max}=0.5$ }}
\label{mode_power}
\end{figure}

\begin{figure}[!t]
\centering
\includegraphics[width=0.3\textwidth]{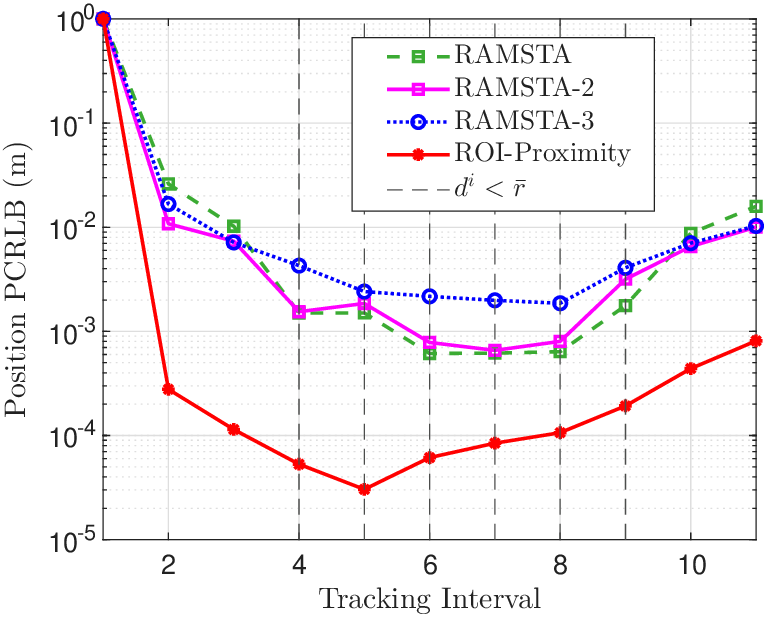}
\caption{\footnotesize{Target tracking performance of the proposed RAMSTA framework under varying benchmarks ($\delta_{\text{min}}=0.2$, $\delta_{\text{max}}=0.8 $ )}}
\label{bench}
\end{figure}

In Fig. \ref{in_out2} we draw a contrast between the position PCRLB and communication sum rate under varying ROI sensitivity thresholds. First we consider the case where the target is outside the  ROI, $i=3$, in Fig. \ref{in_out2}-i . Here, as the distance between the target's position and the center of the ROI increases, $\delta^i$ approaches $\delta_{\text{min}}$ (\ref{trade_off}). Thus, outside the ROI, decreasing $\delta_{\text{min}}$ whiles increasing $\delta_{\text{max}}$ shifts the communication-sensing trade-off in favor of communication performance resulting in an increase in sum rate and position PCRLB. Specifically, when $\delta_{\text{max}}=0.8$ and $\delta_{\text{min}}=0.2$, we observe the highest sum rate of $278.9$ Mbps but and the worst position PCRLB of $0.0099$ m. Conversely, a sum rate and position PCRLB of $260.9$ Mbps and $0.0032$ m are respectively observed when $\delta_{\text{max}}=0.7$ and $\delta_{\text{min}}=0.3$. This accounts for about a $6.5$\% decrease and a 67.7\% increase in communication and sensing performance respectively. $\delta_{\text{max}}=\delta_{\text{min}}=0.5$ weights communication and sensing equally and removes the impact of the ROI tuning parameter. The converse occurs inside the ROI as observed in Fig. \ref{in_out2}-ii. The lowest position PCRLB and sum rate occurs at  $\delta_{\text{max}}=0.8$, $\delta_{\text{min}}=0.2$ with $0.0006$ m and $91$ Mbps respectively. Furthermore, when $\delta_{\text{max}}=0.7$, $\delta_{\text{min}}=0.3$, the position PCRLB and sum rate become $0.0007$ m and $116.7$ Mbps respectively, accounting for an increase in communication and a decrease in sensing performance. 

Fig. \ref{mode_power} shows the mode selection and task assignment results for $\delta_{min}=0.2$, $\delta_{max}=0.8$  and $\delta_{min}=\delta_{max}=0.5$ in Fig. \ref{mode_power}a and Fig. \ref{mode_power}b respectively. In Fig. \ref{mode_power}a it is clear that when $\delta_{\text{min}}=0.2$, $\delta_{\text{max}}=0.8$, the higher communication priority outside the ROI ($i=2,3,10,11$) allows a higher number of transmit APs all of which perform at least a communication task. This represents a favorable mode selection to yield the benefits of distributed MIMO and thus improved communication performance.
Conversely, inside the ROI a fair balance between transmit ISAC and receive APs is maintained to yield a sufficient number of bistatic links to improve diversity and hence sensing performance. At $\delta_{min}=\delta_{max}=0.5$, Fig. \ref{mode_power}b, a balance between communication and sensing favorable mode and task assignment is maintained across tracking intervals. 

In Fig. \ref{bench}, we compare the position PCRLB performance of the RAMSTA framework against different benchmarks:
    \begin{itemize}
    \item \textbf{RAMSTA-2}: Here an ROI-unaware equal power and subcarrier allocation is assigned via a similar heuristic approach in Algorithm \ref{alg:cap2}.
    \item \textbf{RAMSTA-3}: The impact of the ROI-aware tuning parameters is isolated to the objective function by setting $\upsilon=J-1$ in (\ref{task4}) paired with a similar power and subcarrier allocation in RAMSTA-2.
    \item \textbf{ROI-Proximity} : A sensing-biased benchmark, which selects a fixed AP mode with the minimum average distance to the ROI center. This results in $1$ receive AP inside the ROI at $(221.9,105.7,7.2)$ m and $5$ transmit APs. Here all transmit APs are assigned ISAC tasks whereas power and subcarrier are allocated as in RAMSTA-2.
\end{itemize} 
 We observe that, the proposed RAMSTA framework successfully maintains a relatively good inside-ROI tracking performance with the minimum of $0.0006$ m occurring at $i=6$.  RAMSTA-2 and RAMSTA-3 on the other hand obtain position PCRLBs of $0.0008$m and $0.0022$m at $i=6$ respectively. The $0.0002$ m performance gain by RAMSTA over RAMSTA-2 is attributed to the fact that RAMSTA-2 is devoid of ROI sensitive power and bandwidth gains inside the ROI. Thus an additional ROI-aware power and  subcarrier allocation can provide a maximum of 25\% performance improvement inside the ROI. However, this comes at the cost of degraded communication and sensing performance inside and outside the ROI respectively.
 On the other hand, RAMSTA-3 imposes less restriction on the maximum number AP assigned per DL UE thus the likelihood of communication only assignment increases. This results in a  $0.0016$ m performance difference compared to RAMSTA at $i=6$. Lastly, the ROI-proximity guarantees a high sensing SNR due to its minimal radar pathloss inside the ROI, resulting in the lowest position PCRLB performance ($0.039$ mm at $i=5$). 
\section{Conclusion}
In this work we  proposed a RAMSTA framework where we design a sigmoid-based ROI proximity parameter to tune the communication-sensing trade-off in a weighted position PCRLB and sum rate minimization problem. The resulting MINLP is solved via a penalized convex-concave procedure. We show that ROI sensitivity thresholds allows the adaptive variation of communication-sensing trade-off with respect to the predicted location of the target and relative to a predefined ROI. Future work would focus on end-to-end resource allocation and beamforming design under polyhedron-shaped ROI's as well as the impact of ROI proximity prediction errors.

{\appendix[Derivation of (\ref{data_f})]\label{sec:AppendixA}}
Let $\mathbf{y} = [\mathbf{y}_1 ^T, \mathbf{y}_2 ^T,\cdots, \mathbf{y}_J ^T  ]^T \in \mathcal{C}^{J^2L_rMN\times 1}$, $\mathbf{y}_r = [\mathbf{y}_{r,1}(1,1) ^T, \cdots, \mathbf{y}_{r,j}(m,n) ^T,\cdots, \mathbf{y}_{r,J}(M,N) ^T  ]^T \in \mathcal{C}^{JL_rMN\times 1}$ be (\ref{rad}), stacked over all subcarriers and transmit-receiver AP pairs such that $\mathbf{y} \sim \mathcal{CN}(\bm{\mu},\bm{\sigma})$. Then the unknown parameter vector is   $\bm{\gamma} = [\bm{\varphi},\bm{\alpha}]$, such that $\bm{\varphi} = [\bm{\varphi}_{1},\cdots,\bm{\varphi}_{r},\cdots,\bm{\varphi}_{J}]$, $\bm{\varphi}_r = [\bm{\varphi}_{r,1},\cdots,\bm{\varphi}_{r,J}]$, $\bm{\varphi}_{r,j} = [f_{r,j}, \tau_{r,j}, \theta_r, \phi_r, \theta_j, \phi_j ]$, $\bm{\alpha} = [\bm{\alpha}_{1},\cdots,\bm{\alpha}_{r},\cdots, \bm{\alpha}_{J} ]$ , and $\bm{\alpha}_r = [\alpha_{r,1},\cdots, \alpha_{r,2},\cdots \alpha_{r,J} ]$. Also let $\mathbf{u}=[x^i, y^i, z^i, \dot{x}^i, \dot{y}^i, \dot{z}^i, \bm{\alpha}]^T $. Then the FIM has the form, \cite{10.5555/151045}
 \begin{align}
 \label{fim2}
     \bar{\mathbf{J}}(\bm{\gamma}) &= 2\mathbb{R}\Big[\Big(  \frac{\partial \bm{\mu}}{\partial \bm{\gamma}} \Big)^H\bm{\sigma}^{-1} \frac{\partial \bm{\mu}}{\partial \bm{\gamma}}\Big] \\
      \label{fim}
        \bar{\mathbf{J}}(\mathbf{u})&= \frac{\partial \bm{\varphi} }{\partial \bm{u}} \bar{\mathbf{J}}(\bm{\gamma})\frac{\partial^T \bm{\varphi} }{\partial \bm{u}} 
 \end{align}
Given (\ref{fim}), $ \frac{\partial \bm{\mu} }{\partial \bm{\gamma}} = \big[ \frac{\partial \bm{\mu}_1^T }{\partial \bm{\gamma}} \cdots\frac{\partial \bm{\mu}_r^T }{\partial \bm{\gamma}} \cdots  \frac{\partial \bm{\mu}_J^T }{\partial \bm{\gamma}} \big]^T$, where 
  \begin{equation}
  \label{app1}
     \frac{\partial \bm{\mu}_r }{\partial \bm{\gamma}} = 
     \begin{bmatrix}
     \frac{\partial \bm{\mu}_{r,1}(1,1) }{\partial \bm{\varphi}_1}  & \cdots & \frac{\partial \bm{\mu}_{r,1}(1,1) }{\partial \bm{\varphi}_J}  & \frac{\partial \bm{\mu}_{r,1}(1,1) }{\partial \bm{\alpha}} \\
         
        \vdots  & \cdots & \vdots & \vdots \\
          \frac{\partial \bm{\mu}_{r,j}(m,n) }{\partial \bm{\varphi}_1} & \cdots & \frac{\partial \bm{\mu}_{r,j}(m,n) }{\partial \bm{\varphi}_J} & \frac{\partial \bm{\mu}_{r,j}(m,n) }{\partial \bm{\alpha}} \\
          \vdots  & \cdots & \vdots & \vdots \\
           \frac{\partial \bm{\mu}_{r,J}(M,N) }{\partial \bm{\varphi}_1}  & \cdots & \frac{\partial \bm{\mu}_{r,J}(M,N) }{\partial \bm{\varphi}_J} & \frac{\partial \bm{\mu}_{r,J}(M,N) }{\partial \bm{\alpha}} 
     \end{bmatrix}
\end{equation}
In (\ref{app1}), $\frac{\partial \bm{\mu}_{r,j}(m,n) }{\partial \bm{\varphi}_r} = [\frac{\partial \bm{\mu}_{r,j}(m,n) }{\partial \bm{\varphi}_{r,1}} \cdots \frac{\partial \bm{\mu}_{r,j}(m,n) }{\partial \bm{\varphi}_{r,j}} \cdots \frac{\partial \bm{\mu}_{r,j}(m,n) }{\partial \bm{\varphi}_{r,J}} ] $. $\frac{\partial \bm{\mu}_{r,j}(m,n) }{\partial \bm{\varphi}_{r,j}} $ can have take four different forms: $\frac{\partial \bm{\mu}_{r,j}(m,n) }{\partial \bm{\varphi}_{r,j}}  = [\frac{\partial \bm{\mu}_{r,j}(m,n) }{\partial f_{r,j}} \frac{\partial \bm{\mu}_{r,j}(m,n) }{\partial \tau_{r,j}} \frac{\partial \bm{\mu}_{r,j}(m,n) }{\partial \theta_{r}} \frac{\partial \bm{\mu}_{r,j}(m,n) }{\partial \phi_{r}} \frac{\partial \bm{\mu}_{r,j}(m,n) }{\partial \theta_{j}} \frac{\partial \bm{\mu}_{r,j}(m,n) }{\partial \phi_{j}}]$, $\frac{\partial \bm{\mu}_{r,j}(m,n) }{\partial \bm{\varphi}_{r,j'}}  = [0, 0,\frac{\partial \bm{\mu}_{r,j}(m,n) }{\partial \theta_{r}} \frac{\partial \bm{\mu}_{r,j}(m,n) }{\partial \phi_{r}}, 0, 0]$, $\frac{\partial \bm{\mu}_{r,j}(m,n) }{\partial \bm{\varphi}_{r',j}}  = [0,0,0,0, \frac{\partial \bm{\mu}_{r,j}(m,n) }{\partial \theta_{j}} \frac{\partial \bm{\mu}_{r,j}(m,n) }{\partial \phi_{j}}]$, and $\frac{\partial \bm{\mu}_{r,j}(m,n) }{\partial \bm{\varphi}_{r',j'}}  = [0,0,0,0, 0, 0]$, where $r'\neq r$ and $j'\neq j$. 
 Let $\frac{\partial \bm{\varphi} }{\partial \bm{u}} = \big[ \frac{\partial \bm{\varphi}_1 }{\partial \bm{u}} \cdots \frac{\partial \bm{\varphi}_J }{\partial \bm{u}}  \frac{\partial \bm{\alpha} }{\partial \bm{u}} \big]$, then given (\ref{app1})  and ignoring terms related to $\alpha$, (\ref{fim}) has the form
\begin{equation}
\label{app2}
    \mathbf{\bar{J}(u)} = \sum_{n=1}^N \sum_{m=1}^M \sum_{r=1}^J \sum_{j=1}^J \sigma_{r,j}^{-1} \begin{bmatrix}
\mathbf{\bar{J}(u)}_p & \mathbf{\bar{J}(u)}_{pv} \\
\mathbf{\bar{J}(u)}_{vp} & \mathbf{\bar{J}(u)}_v 
\end{bmatrix}
\end{equation}
where $\mathbf{\bar{J}(u)}_p$, $\mathbf{\bar{J}(u)}_v$ characterize the FIM of position and velocity respectively, whereas the off diagonal elements characterize their coupling. Furthermore,
\begin{equation}
    \mathbf{\bar{J}(u)}_p = \sum_{q \in \varepsilon} \sum_{p \in \varepsilon} \begin{bmatrix}
\frac{\partial q}{\partial x} \Xi\frac{\partial q}{\partial x} & \frac{\partial q}{\partial x} \Xi\frac{\partial q}{\partial y}& \frac{\partial q}{\partial x} \Xi\frac{\partial q}{\partial z}\\
\frac{\partial q}{\partial y} \Xi\frac{\partial q}{\partial x}& \frac{\partial q}{\partial y} \Xi\frac{\partial q}{\partial y}  & \frac{\partial q}{\partial y} \Xi\frac{\partial q}{\partial z} \\
\frac{\partial q}{\partial z} \Xi\frac{\partial q}{\partial x}& \frac{\partial q}{\partial z} \Xi\frac{\partial q}{\partial y} & \frac{\partial q}{\partial z} \Xi\frac{\partial q}{\partial z}
\end{bmatrix}
\end{equation}
where $ \varepsilon=\{f_{r,j}, \tau_{r,j}, \theta_r, \phi_r,\theta_j, \phi_j \}$ and $\Xi = 2\mathbb{R}\Big\{\frac{\partial \bm{\mu}_{r,j}(m,n)}{\partial q}^H \frac{\partial \bm{\mu}_{r,j}(m,n)}{\partial p} \Big\}$. Also,
\begin{equation}
    \mathbf{\mathbf{\bar{J}(u)}}_v = \begin{bmatrix}
 \frac{\partial f_{r,j}}{\partial \dot{x}}\Lambda\frac{\partial f_{r,j}}{\partial \dot{x}} & \frac{\partial f_{r,j}}{\partial \dot{x}}\Lambda\frac{\partial f_{r,j}}{\partial \dot{y}}& \frac{\partial f_{r,j}}{\partial \dot{x}}\Lambda\frac{\partial f_{r,j}}{\partial \dot{z}}\\
\frac{\partial f_{r,j}}{\partial \dot{y}}\Lambda\frac{\partial f_{r,j}}{\partial \dot{x}}&\frac{\partial f_{r,j}}{\partial \dot{y}}\Lambda\frac{\partial f_{r,j}}{\partial \dot{y}}  &\frac{\partial f_{r,j}}{\partial \dot{y}}\Lambda\frac{\partial f_{r,j}}{\partial \dot{z}} \\
\frac{\partial f_{r,j}}{\partial \dot{z}}\Lambda\frac{\partial f_{r,j}}{\partial \dot{x}}& \frac{\partial f_{r,j}}{\partial \dot{z}}\Lambda\frac{\partial f_{r,j}}{\partial \dot{y}}& \frac{\partial f_{r,j}}{\partial \dot{z}}\Lambda\frac{\partial f_{r,j}}{\partial \dot{z}}
\end{bmatrix}
\end{equation}
where $\Lambda = 2\mathbb{R}\Big\{\frac{\partial \bm{\mu}_{r,j}(m,n)}{\partial f_{r,j}}^H \frac{\partial \bm{\mu}_{r,j}(m,n)}{\partial f_{r,j}} \Big\} $. (\ref{data_f}) is obtained by factorizing $(1-a_r)a_jp_{j}^i(m)\bar{a}_{j,0}a_j\eta_{j}^i(m)$, where $\tilde{\mathbf{J}}(\mathbf{u})$ is the remainder after factorization. The derivation of the off-diagonal terms of (\ref{app2}) and the derivatives are omitted for brevity.

The measurement noise covariance is similarly obtained by $\mathbf{R}_{r,j} = \tilde{\mathbf{J}}(\bm{\gamma})^{-1}$, where $\tilde{\mathbf{J}}(\bm{\gamma})^{-1}$ is the resulting form of (\ref{fim2}) when $\bm{\varphi}_{r,j} = [f_{r,j}, \tau_{r,j}, \theta_r, \phi_r ]$.

\section*{Acknowledgment}
The work is sponsored by Dutch 6G Future Network Services.



%
\bibliographystyle{ieeetr}

\bibliography{EuCnC}

@article{xiong_coalition_2023,
	title = {Coalition Game of Radar Network for Multitarget Tracking via Model-Based Multiagent Reinforcement Learning},
	volume = {59},
	issn = {1557-9603},
	url = {},
	doi = {10.1109/TAES.2022.3208865},
	pages = {2123--2140},
	number = {3},
	journal = {{IEEE} Trans. Aerosp. Electron. Syst.},
	author = {Xiong, Kui and others},
	urldate = {2025-04-09},
    month = {Jun.},
	date = {2023-06},
    year = {2023},
}

@ARTICLE{9540344,
  author={Zhang, J. Andrew and others},
  journal={IEEE J. Sel. Areas Commun.}, 
  title={An Overview of Signal Processing Techniques for Joint Communication and Radar Sensing}, 
  year={2021},
  volume={15},
  number={6},
  pages={1295-1315},
  doi={10.1109/JSTSP.2021.3113120}}

@INPROCEEDINGS{10571110,
  author={Liu, Sifan and others},
  booktitle={Proc. 2024 IEEE WCNC}, 
  title={Cooperative Cell-Free ISAC Networks: Joint BS Mode Selection and Beamforming Design}, 
  year={2024},
  volume={},
  number={},
  pages={1-6},
  doi={10.1109/WCNC57260.2024.10571110}}

@book{10.5555/151045,
author = {Kay, Steven M.},
title = {Fundamentals of Statistical Signal Processing: Estimation Theory},
year = {1993},
isbn = {0133457117},
publisher = {Prentice-Hall, Inc.},
address = {USA}
}

@ARTICLE{resource,
  author={Zhong, Chen and others},
  journal={IEEE Trans. Veh. Technol.}, 
  title={Resource Allocation for Dynamic {TDD}-Enabled Integrated Sensing and Communication Systems}, 
  year={2025},
  volume={74},
  number={6},
  month = {Jun.},
  pages={9357-9369},
  doi={10.1109/TVT.2025.3541268}}

@ARTICLE{motion_model,
  author={Rong Li, X. and Jilkov, V.P.},
  journal={IEEE Trans. Aerosp. Electron. Syst.}, 
  title={Survey of maneuvering target tracking. Part I. Dynamic models}, 
  year={2003},
  volume={39},
  number={4},
  pages={1333-1364},
  doi={10.1109/TAES.2003.1261132}}

@book{rad_range,
author = {Mark A. Richards  and James A. Scheer  and William A. Holm },
title = {Principles of Modern Radar: Basic principles},
publisher = {The Institution of Engineering and Technology},
year = {2010},
doi = {10.1049/SBRA021E},
edition   = {},
URL = {https://digital-library.theiet.org/doi/abs/10.1049/SBRA021E},
eprint = {https://digital-library.theiet.org/doi/pdf/10.1049/SBRA021E}
}

@ARTICLE{steer,
  author={Zhao, Chuanbin and Feng, Yuan and Luo, Hongliang and Gao, Feifei and Liu, Fan and Jin, Shi},
  journal={IEEE Trans. Wireless Commun.}, 
  title={Networked ISAC-Based UAV Tracking and Handover Toward Low-Altitude Economy}, 
  year={2025},
  volume={24},
  number={9},
  pages={7670-7685},
  doi={10.1109/TWC.2025.3562396}}

@ARTICLE{pcrlb,
  author={Tichavsky, P. and Muravchik, C.H. and Nehorai, A.},
  journal={IEEE Trans. Signal Process.}, 
  title={Posterior Cramer-Rao bounds for discrete-time nonlinear filtering}, 
  year={1998},
  volume={46},
  number={5},
  pages={1386-1396},
  doi={10.1109/78.668800}}

@INPROCEEDINGS{bilinear,
  author={Vu, Quang-Doanh and Nguyen, Kien-Giang and Juntti, Markku},
  booktitle={2016 IEEE Global Communications Conference (GLOBECOM)}, 
  title={Max-Min Fairness for Multicast Multigroup Multicell Transmission under Backhaul Constraints}, 
  year={2016},
  volume={},
  number={},
  pages={1-6},
  doi={10.1109/GLOCOM.2016.7841981}}

@article{mccormick_computability_1976,
	title = {Computability of global solutions to factorable nonconvex programs: Part I — Convex underestimating problems},
	volume = {10},
	issn = {1436-4646},
	url = {https://doi.org/10.1007/BF01580665},
	doi = {10.1007/BF01580665},
	pages = {147--175},
	number = {1},
	journaltitle = {Mathematical Programming},
	author = {{McCormick}, Garth P.},
	date = {1976-12-01},
}

@ARTICLE{sdp,
  author={Xie, Mingchi and Yi, Wei and Kirubarajan, Thia and Kong, Lingjiang},
  journal={IEEE Trans. Signal Process.}, 
  title={Joint Node Selection and Power Allocation Strategy for Multitarget Tracking in Decentralized Radar Networks}, 
  year={2018},
  volume={66},
  number={3},
  pages={729-743},
  doi={10.1109/TSP.2017.2777394}}

@ARTICLE{multistatic,
  author={Behdad, Zinat and others},
  journal={IEEE Trans. Wireless Commun.}, 
  title={Multi-Static Target Detection and Power Allocation for Integrated Sensing and Communication in Cell-Free Massive MIMO}, 
  year={2024},
  volume={23},
  number={9},
  pages={11580-11596},
  doi={10.1109/TWC.2024.3383209}}

@ARTICLE{mode3,
  author={Ren, Zhichu and Pan, Cunhua and Ren, Hong and Wang, Dongming and Xu, Lexi and Wang, Jiangzhou},
  journal={IEEE Trans. Wireless Commun.}, 
  title={Two-Timescale Design for AP Mode Selection and Power Allocation of Cooperative ISAC Networks}, 
  year={2026},
  volume={25},
  number={},
  pages={815-831},
  doi={10.1109/TWC.2025.3587068}}

@ARTICLE{multistaticss,
  author={Zhou, Quan and Cheng, Ting and Cao, Congchong and He, Sichen and Song, Jiaming and He, Zishu},
  journal={IEEE Internet Things J.}, 
  title={Adaptive Channel Allocation in Multistatic Passive Radar System for Multiple Maneuvering Targets Tracking With Missed Detection}, 
  year={2026},
  volume={13},
  number={5},
  pages={9492-9504},
  doi={10.1109/JIOT.2025.3645608}}

\end{document}